\documentstyle[11pt]{article}

\def\dalemb#1#2{{\vbox{\hrule height .#2pt
        \hbox{\vrule width.#2pt height#1pt \kern#1pt
                \vrule width.#2pt}
        \hrule height.#2pt}}}

\let\a=\alpha \let\b=\beta   \let\e=\epsilon

 \def\bd{\begin{document}} \def\ed{\end{document}}
\def\ds{\documentstyle} \let\fr=\frac \let\bl=\bigl \let\br=\bigr
\let\Br=\Bigr \let\Bl=\Bigl 
\let\bm=\bibitem
\let\na=\nabla
\let\pa=\partial \let\ov=\overline 
\newcommand{\be}{\begin{equation}} 
\newcommand{\ee}{\end{equation}} 
\def\ve{\varepsilon}
\def\ba{\begin{array}}
\def\ea{\end{array}}
\def\ft#1#2{{\textstyle{{\scriptstyle #1}\over {\scriptstyle #2}}}}
\def\fft#1#2{{#1 \over #2}}
\def\del{\partial}
\def\sst#1{{\scriptscriptstyle #1}}
\def\oneone{\rlap 1\mkern4mu{\rm l}}
\def\e7{E_{7(+7)}}
\def\td{\tilde}
\def\bog{Bogomol'nyi\ }
\newcommand{\ho}[1]{$\, ^{#1}$}
\newcommand{\hoch}[1]{$\, ^{#1}$}
\newcommand{\bea}{\begin{eqnarray}} 
\newcommand{\eea}{\end{eqnarray}} 
\newcommand{\ra}{\rightarrow}
\newcommand{\lra}{\longrightarrow}
\newcommand{\Lra}{\Leftrightarrow}
\newcommand{\ap}{\alpha^\prime}
\newcommand{\bp}{\tilde \beta^\prime}
\newcommand{\tr}{{\rm tr} }
\newcommand{\Tr}{{\rm Tr} } 
\newcommand{\NP}{Nucl. Phys. }
\newcommand{\tamphys}{\it Center for Theoretical Physics,
Texas A\&M University, College Station, Texas 77843}

\newcommand{\auth}{H. L\"u and C.N. Pope}

\begin{document}
\begin{flushright}
\hfill{CTP TAMU-22/96}\\
\hfill{SISSA 89/96/EP}\\
\hfill{hep-th/9606047}\\
\end{flushright}

\vspace{20pt}

\begin{center}
{\large {\bf Fission and Fusion Bound States of $p$-brane Solitons}}

\vspace{30pt}

\auth

\vspace{15pt}

{\tamphys}


\vspace{15pt}

{\it SISSA, Via Beirut No. 2-4, 34013 Trieste, Italy }



\vspace{40pt}

\underline{ABSTRACT}
\end{center}

          Supersymmetric $p$-branes that carry a single electric or magnetic
charge and preserve $1/2$ of the supersymmetry have been interpreted as
the constituents from which all supersymmetric $p$-branes can be constructed
as bound states, albeit with zero binding energy.  Here we extend the 
discussion to non-supersymmetric $p$-branes, and argue that they also can
be interpreted as bound states of the same basic supersymmetric constituents.
In general, the binding energy is non-zero, and can be either positive or 
negative depending on the specific choice of constituents.   In particular,
we find that the $a=0$ Reissner-Nordstr\o{m} black hole in $D=4$ can be built
from different sets of constituents such that it has zero, positive or negative
binding energy.

{\vfill\leftline{}\vfill
\vskip	10pt
\footnoterule
{\footnotesize
	Research supported in part by DOE 
Grant DE-FG05-91-ER40633 and \vskip	-12pt}  \vskip	10pt
{\footnotesize 
      EC Human Capital and Mobility Programme under contract ERBCHBGCT920176.  
} 
}

\pagebreak
\setcounter{page}{1}

\section{Introduction}

      Isotropic $p$-brane solitons in M-theory or string theory have been
extensively studied, and their classification has been discussed from
various points of view.  One approach is to organise the various 
solutions using the U duality group of the theory.  In particular, it
was shown that $p$-brane solutions form representations of the U Weyl
group \cite{lpsweyl}.  Another approach is to interpret lower dimensional 
solutions from the viewpoint of the fundamental dimension of the theory, 
namely $D=11$ in the case of M-theory.  In other words, the 
lower-dimensional solutions can be oxidised, by the inverse of the 
Kaluza-Klein reduction procedure, to solutions in $D=11$.  It has been shown 
that lower dimensional supersymmetric $p$-branes can be viewed as 
intersecting M-branes [2-7] or boosted intersecting M-branes \cite{kklp} in 
$D=11$.  A third approach, which until now has also been applied only to 
supersymmetric $p$-branes, is to view those solutions that carry more than 
one kind of charge as bound states of single-charge solutions 
\cite{r,kklp,dr}. For example, the $a=1, 1/\sqrt3$ and 0 black holes in $D=4$ 
can be viewed as bound states of two, three or four $a=\sqrt3$ black holes 
\cite{r}.  Another example is provided by the dyonic string \cite{dfkr} in 
$D=6$, which can be viewed as a bound state of an electric and a magnetic 
string \cite{kklp}.  All the above solutions Strictly speaking,
the term bound state is a misnomer, since the binding energy is actually
zero for these supersymmetric $p$-branes.  This zero binding energy is
consistent with the fact that the charges in the above multi-charge solutions 
can be located independently; the bound states can be "pulled apart" into
constituents that can sit in static equilibrium at any separation.  This
type of multi-charge, multi-center solution was first discussed in 
\cite{klopv}, where a four-charge supersymmetric black hole in $D=4$ was
``split'' into two $a=1$ two-charge black holes (which themselves can 
be further split into $a=\sqrt3$ black holes). 

       The majority of $p$-brane solutions are in fact not supersymmetric
\cite{lpsol}.  The purpose of this paper is to extend the previous discussion 
of bound states to encompass these non-supersymmetric solutions.  As we shall
see, all these solutions can also be viewed as bound states of the same 
supersymmetric building blocks, namely the single-charge $p$-branes.  The
major difference from the supersymmetric bound states is that now the binding
energy is non-zero.   In some cases, the binding energy is positive, implying
that the supersymmetric building blocks will undergo a spontaneous fusion
to form the non-supersymmetric $p$-brane.  In other cases, the binding
energy is negative, and the non-supersymmetric $p$-brane will undergo a 
spontaneous fission into its supersymmetric constituents.

       The binding energy of a $p$-brane can easily be calculated by 
comparing its mass with the sum of the masses of its individual constituents
when their locations are widely separated.  Of course if the binding energy
is non-zero, this configuration will not be an exact solution.  However, it
can be made arbitrarily good by taking the separations to be sufficiently
large.  We shall discuss the binding energy for single-scalar solutions
in section 2.   In these solutions, the charges carried by the various
participating field strengths arise in fixed ratios, which implies
that they are formed as bound states of constituents whose charges have the
same ratios.  In the case of supersymmetric $p$-branes, more general solutions
are known in which the charges can be independently specified \cite{lpmulti}, 
implying that
these are bound states of constituents with independent charges.   The
analogous solutions with independent charges are not in general known for
the non-supersymmetric cases, and so the exact discussion is generally 
restricted to the cases where the constituents have the necessary fixed
ratios of charges.  In section 3, however, we shall present an explicit
exact solution with two independent charges in one particular 
non-supersymmetric case, namely a black hole dyon in $D=4$, with one field
strength that carries both an electric charge $Q_e$ and a magnetic charge 
$Q_m$.  The mass of the bound state is
\be
m=(Q_e^{2/3} + Q_m^{2/3})^{3/2} \ ,\label{dyonmass}
\ee
while the widely-separated electric and magnetic black hole constituents have
total mass $m_\infty = Q_e+Q_m$.  Thus the binding energy in this case is 
negative.
If the charges $Q_e$ and $Q_m$ are equal, the solution reduces to an 
extremal $a=0$
Reissner-Nordstr\o{m} black hole.  This is quite distinct from the usual
four-charge $a=0$ Reissner-Nordstr\o{m} black hole which, being 
supersymmetric, has zero binding energy.  By contrast, this new two-charge
Reissner-Nordstr\o{m} black hole is non-supersymmetric, and is like 
a dyon fission bomb with a yield of about  $29\%$ of the mass of the dyon.  
In fact we can also construct an eight-charge
$a=0$ Reissner-Nordstr\o{m} black hole with positive binding energy.

	We have not found exact solutions for any other non-supersymmetric
$p$-branes with independent charges.  Instead, in section 4, we present a 
general perturbative analysis for charges which are close to the fixed 
ratios of the exact single-scalar solutions. In certain cases, we can use 
these results to conjecture the analogue of the mass formula (\ref{dyonmass}).

\section{Binding energy of single-scalar $p$-branes}

    The Kaluza-Klein reduction of eleven-dimensional supergravity to $D$
dimensions gives rise to various field strengths coming both from the 
4-form and from the vielbein in $D=11$.  Denoting the compactified $(11-D)$
internal indices by $i, j, \cdots$, we have $F_4$, $F_3^{(i)}$, $F_2^{(ij)}$
and $F_1^{(ijk)}$ from the former, and ${\cal F}^{(i)}_2$ and
${\cal F}_1^{(ij)}$ from the latter, where the subscript index denotes the 
degree of the form.  The detailed expression for the bosonic Lagrangian can
be found in \cite{lpsol}.  It admits a consistent truncation to the Lagrangian
\be
e^{-1}{\cal L}= R -\ft12 (\del\vec\phi)^2 -\fft{1}{2 n!} \sum_{\a=1}^N
e^{-\vec a_\a\cdot \vec\phi} \, F_\a^2 + {\cal L}_{\sst{FFA}}\ ,
\label{dlag}
\ee
where $\vec\phi=(\phi_1,\cdots,\phi_{\sst{N}})$, $F_\a$ are a set of $N$
antisymmetric tensor field strengths of rank $n$, and $\vec a_\a$ are 
the associated constant vectors characterising their couplings to the 
dilatonic scalars $\vec\phi$.  The term ${\cal L}_{\sst{FFA}}$ represents
the dimensional reduction of the $FFA$ term in $D=11$.  This, together
with the Chern-Simons modifications to the field strengths, vanishes for
the solutions we shall construct, and we shall not consider them further.  
(This puts constraints on the possible
charges that can be used to construct $p$-brane solutions; a 
full discussion of these constraints can be found in \cite{lpsol}.)
The $n$-rank field strengths can be used to construct elementary $p$-branes
with world-volume dimension $d=n-1$, or solitonic $p$-branes with $d=D-n-1$.
In both cases, the $p$-brane metric takes the form
\be
ds^2 = e^{2A} dx^\mu dx^\nu \eta_{\mu\nu} + e^{2B} dy^m dy^m\ ,
\ee
where $x^\mu$ are the coordinates on the $d$-dimensional world volume.  The
functions $A$ and $B$ depend only on the coordinates $y^m$ of the transverse
space.  In all the extremal solutions that we shall be considering in this
paper, they satisfy the relation $dA+\td dB=0$, where $\td d\equiv D-d-2$.
     
     A further truncation to the single-scalar Lagrangian
\be
e^{-1} {\cal L}= R -\ft12 (\del\phi)^2 -\fft{1}{2 n!} \, e^{-a\phi}\,
F^2 \label{ss}
\ee
is possible, where $a$, $\phi$ and $F$ are given by \cite{lpsol}
\bea
a^2 &=& \Big(\sum_{\a ,\b} (M^{-1})_{\a\b}\Big)^{-1}\ ,\qquad 
\phi=a \sum_{\a ,\b} (M^{-1})_{\a\b} \, \vec a_\a\cdot \vec\phi \ ,
\nonumber\\
F_\a^2 &=& a^2 \sum_\b (M^{-1})_{\a\b} \, F^2\ ,\label{cons}
\eea
and $M_{\a\b}=\vec a_\a \cdot \vec a_\b$.  The parameter $a$ can be
conveniently re-expressed as
\be
a^2=\Delta - \fft{2 d\td d}{D-2}\ ,\label{delta}
\ee
where $\Delta$ is preserved under dimensional reduction \cite{lpstain}. 
The equations of motion following from (\ref{ss})
admit extremal $p$-brane solutions given by
\bea
ds^2 &=& \Big(1+\fft{k}{r^{\td d}}\Big)^{-\ft{4\td d}{\Delta (D-2)}}\,
dx^\mu dx^\nu \eta_{\mu\nu} + \Big(1+\fft{k}{r^{\td d}} \Big)^{\ft{4 d}{
\Delta (D-2)}} dy^m dy^m \ ,\nonumber\\
e^\phi &=& \Big(1+\fft{k}{r^{\td d}}\Big)^{\ft{2a}{\epsilon \Delta}}\ ,
\label{ssol}
\eea
where $\epsilon=1$ for elementary solutions and $\epsilon=-1$ for solitonic
solutions.   The field strength $F$ carries electric or magnetic charge $Q$,
$k=2\sqrt{\Delta} Q/\td d$, and the mass per unit $p$-volume is given by
\be
m=\fft{2Q}{\sqrt\Delta}\ .\label{ssmass}
\ee

      The mass (\ref{ssmass}) of a single-scalar $p$-brane solution can be
compared with the total mass $m_\infty$ of its widely-separated constituents.  
If follows from (\ref{cons}) that each field strength $F_\a$ carries a charge
\be
Q_\a  \equiv  Q\, c_\a \ ,\qquad {\rm with}\qquad
c_\a\equiv a\Big(\sum_\b (M^{-1})_{\a\b}\Big)^{1/2}\ .
\label{sscharge}
\ee
The constants $c_\a$ describe the fixed fractions of the total normalised 
charge $Q$ that are carried by the field strengths $F_\a$.  Note that 
$\sum_\a c_\a^2=1$.  The total mass $m_\infty$ is 
\be
m_\infty = \sum_\a Q_\a = Q \sum_\a c_\a\ ,
\ee
since each constituent is a $\Delta=4$ solution.    The binding energy
is then given by
\be
\delta E  = Q \Big( \sum_\a c_\a - \fft{2}{\sqrt{\Delta}} 
\Big)\ .\label{ssbe}
\ee

      For supersymmetric $p$-branes, the dilaton vectors $\vec a_\a$ satisfy
the dot products
\be
M_{\a\beta} = 4 \delta_{\a\beta} - \fft{2 d \td d}{D-2}\ ,\label{susydot}
\ee
which implies that the $c_\a$ are all equal, and that $\Delta=4/N$.  Thus 
from (\ref{ssbe}) we recover the previously-known result that the 
supersymmetric $p$-branes have zero binding energy.

      For non-supersymmetric solutions, the binding energy can be of either
sign, depending on the detailed structure of the dot products of the dilaton
vectors.   The discussion becomes particularly simple for $p$-brane solutions
using 3-form field strengths ({\it i.e.}\ elementary strings or solitonic
$(D-5)$-branes). 
The dot products $M_{\a\beta}$ in this case are given by \cite{lpsol}
\be
M_{\a\beta} = 2 \delta_{\a\beta} - \fft{2(D-6)}{D-2}\ ,
\ee
implying that the $c_\a$ are again all equal, $c_\a=1/\sqrt N$, and 
$\Delta=2 + 2/N$.  Thus the binding energy (\ref{ssbe}) is
\be
\delta E = Q\sqrt{N} \Big( 1 - \sqrt{\ft{2}{N+1}} \Big)\ ,\label{3fssbe}
\ee
which is positive for all $N\ge 2$.  Thus it is energetically favourable
for a set of $N$ widely-separated $\Delta= 4$ constituents carrying the
appropriate set of charges to coalesce to form
a $p$-brane soliton of this kind.  The exact solution that would describe this
collapse would of course be non-static and extremely complicated.  It would
also be non-supersymmetric; however, in the limit where the 
separation between the constituents goes to infinity, supersymmetry would be
asymptotically restored locally in the neighbourhood of each constituent.  
A similar asymptotic local enhancement of supersymmetry also occurs for the
supersymmetric bound states.  The difference in that case however is that
the supersymmetry is never totally broken even when the constituents 
coalesce.  Furthermore the configuration is static, and hence it is somewhat 
misleading to refer to the supersymmetric $p$-branes as bound states, since
the constituents will remain in neutral equilibrium at any separation.
On the other hand, the non-supersymmetric $p$-branes are consequences of the
natural evolution of widely-separated supersymmetric constituents.\footnote{
It is worth remarking here that all the extremal single-scalar
$p$-branes admit multi-center generalisations, implying a non-force
condition, whether or not they are
supersymmetric \cite{lpsdr}.  However, the supersymmetric ones allow a much 
more 
general kind of separation of centers, in which charges of different species
can be located independently in the transverse space.  Obviously, such
static solutions with separated charge species are not possible for the 
non-supersymmetric $p$-branes, precisely because their binding energy is
non-zero.} Note  that in addition to 3-form solutions in M-theory, 
the above solutions can also be used to describe  
a $\Delta = 3$ string with vanishing dilaton in type IIB supergravity, 
which involves both the NS-NS and R-R 3-form field strengths. 

	In the above supersymmetric and non-supersymmetric examples,
we have $c_\a$'s that are all equal.  It follows from (\ref{sscharge}) that
each individual field strength carries an equal charge.  Further 
non-supersymmetric equal-charge $p$-branes can also be constructed using 
2-form field strengths.  For example, the dot products of the dilaton
vectors of the 2-form field strengths ${\cal F}_\a$ coming from the
vielbein are given by
\be
M_{\a\beta} = 2 \delta_{\a\beta} + \fft{2}{D-2}\ .
\ee
it is straightforward to verify that $c_\a$ are equal, and that $\Delta=2 +
2/N$.  Thus the binding energy for elementary black holes or solitonic
$(D-4)$-branes carrying equal charges of these kinds is again given 
by (\ref{3fssbe}).

       While all supersymmetric single-scalar solutions carry equal charges, 
in many non-supersymmetric cases the ratios of the individual charges can be 
different.  This can occur when a solution involves more than two different
type of charges.   A simple example is provided by an $a=0$ black hole
in $D=9$.   The solution involves all the three 2-form field strengths, namely
${\cal F}^{(1)}$ and ${\cal F}^{(2)}$ coming from the $D=11$ vielbein and
$F^{(12)}$ coming from the 4-form in $D=11$.\footnote{This is the first of 
a number of examples of non-supersymmetric $a=0$ black holes in $D<9$ that
lie outside the classification given in \cite{lpsol}, involving $(12-D)$
2-form field strengths.} It has $\Delta = 12/7$
and the fractions $c_\a$ of the total normalised charge $Q$ carried by each 
field strength turn out to be $\sqrt{2/7}$, $\sqrt{2/7}$ and $\sqrt{3/7}$
respectively.   Thus it follows from (\ref{ssbe}) that the 9-dimensional $a=0$
black hole has a positive binding energy.   Another example is an $a=0$ 
eight-charge Reissner-Nordstr\o{m} black hole in $D=4$, which is 
non-supersymmetric.  Indeed it was well known that there exist supersymmetric 
Reissner-Nordstr\o{m} black holes with zero binding energy, which can carry, 
for example, three electric charges and one magnetic charges by the 2-form 
field strengths $F^{(12)}$, $F^{(34)}$, $F^{(56)}$ and ${\cal F}^{(7)}$ 
respectively.  In the non-supersymmetric case we are discussing here, there are
five electric charges carried by the 2-forms ${\cal F}^{(3)}, \cdots, 
{\cal F}^{(7)}$, and three magnetic charges carried by ${\cal F}^{(1)}$,
${\cal F}^{(2)}$ and $F^{(12)}$.  The solution has $\Delta =1$,
and the fractions $c_\a$ are $1/\sqrt{12}$ for each electric charge, and 
$1/\sqrt{6}$, $1/\sqrt{6}$ and $1/2$ for the three magnetic charges.  Thus
the binding energy is again positive.   In the next section, we shall
describe a two-charge dyonic black hole with negative binding energy.

     Let us, at this point, use $D=9$ as an example to illustrate the 
different types of bound-state black holes mentioned above.  There are three
basic black-hole building blocks in $D=9$, namely those where 
${\cal F}^{(1)}$, ${\cal F}^{(2)}$ or $F^{(12)}$ carry the electric charge.
We shall call them type A, B and C respectively.  We can build three
two-charge bound states: AC and BC are both supersymmetric with $\Delta=2$;
in fact they form a doublet under the U Weyl group $S_2$ \cite{lpsweyl}.
AB is non-supersymmetric with $\Delta=3$, and it has positive binding
energy.  Thus if two constituents A and C, or B and C, are initially separated,
they will remain in neutral equilibrium.  On the other hand, if the 
constituents A and B are intitially separated, they will tend to attract
one another.   Similarly, as we mentioned above, there is also an ABC bound 
state with positive binding energy.

     In all the  examples that we have discussed explicitly above, the 
binding energy is non-negative.  This seems to be true for the majority of 
the $p$-brane solutions.  However, there are cases where the binding energy
can be negative.  For example there is a solitonic string in $D=4$, with
charges carried by the 1-form field strengths ${\cal F}^{(12)}$, 
${\cal F}^{(34)}$ and $F^{(135)}$, in the fractions $c_\a = \sqrt{3/10}$,
$\sqrt{3/10}$ and $2/\sqrt{10}$ respectively.  The solution has $\Delta =2/5$,
and hence the binding energy $\delta E= Q(2\sqrt3 - 8)/\sqrt{10}$
is negative.

\section{A black hole dyon in $D=4$}

       In $D=2n$, $n$-form field strengths can carry both electric and
magnetic charges simultaneously, giving rise to dyonic $(n-2)$-branes.
As discussed in \cite{lpsol}, there are two kinds of dyonic $p$-branes. The
first kind involves more than one field strength, with each field strength
carrying either electric or magnetic charge but not both.  An example
is the supersymmetric four-charge Reissner-Nordstr\o{m} black hole, which
we discussed in the previous section.  The second kind is more genuinely 
dyonic, in that each field strength carries both electric and magnetic charge.
In this paper, we reserve the term ``dyon'' exclusively for dyonic $p$-branes 
of the second kind.  An example is the dyonic string in $D=6$ \cite{dfkr}, 
which preserves $1/4$ of the supersymmetry.  Another example is a dyonic 
black hole using two field strengths, 
for example $F^{(12)}$ and ${\cal F}^{(3)}$, with electric charges $Q_1$ 
and $Q_2$, and magnetic charges $Q_2$ and $Q_1$ respectively \cite{lpsol}.  
The second example is non-supersymmetric \cite{lpsol}.   
It reduces to a non-supersymmetric Reissner-Nordstr\o{m} black hole in $D=4$, 
with zero binding energy.  This is understandable since it also involves four 
charges, as in the supersymmetric case.

       In this section, we shall construct a new dyonic black hole in $D=4$,
with only one field strength, carrying both electric charge $Q_e$ and 
magnetic charge $Q_m$.   It is in fact the extremal limit of the dyonic Toda
black hole constructed in \cite{lpxtoda}.  The equations of motion in the 
extremal limit are given by \cite{lpxtoda}
\bea
&&\phi'' = 8\sqrt3 (Q_e^2 e^{\sqrt3 \phi} - Q_m^2 e^{-\sqrt3 \phi}) e^{2A}
\qquad A'' = 4 (Q_e^2 e^{\sqrt3 \phi} + Q_m^2 e^{-\sqrt3\phi}) e^{2A}
\ ,\nonumber\\
&&4A'^2 + \phi'^2 = 16 (Q_e^2 e^{\sqrt3 \phi} + Q_m^2 e^{-\sqrt3 \phi}) e^{2A}
\ ,\label{dyoneq}
\eea
where a prime denotes a derivative with respect to $\rho \equiv 1/r$.  Defining
new functions $q_1$ and $q_2$ by
\be
A = \ft14 \big( q_1 + q_2 - 2\log(16Q_eQ_m)\big)\ ,
\qquad \phi = \ft{\sqrt3}{2} (q_2-q_1) + \ft{1}{\sqrt3} \log\fft{Q_m}{Q_e}
\ ,
\ee
the equations of motion (\ref{dyoneq}) become
\bea
&&q_1'' = e^{2q_1 - q_2}\ ,\qquad q_2'' = e^{2q_2 - q_1}\ ,\label{toda}\\
&&H\equiv \ft13 (p_1^2 + p_2^2 + p_1p_2) - e^{2q_1 - q_2} -
e^{2q_2 - q_1} = 0\ ,\label{hamilton}
\eea
where $H(p_1,p_2,q_1,q_2)$ is the Hamiltonian.  Thus Hamilton's equations 
$q_i' = \del H/\del p_i$ imply that $p_1 = 2q_1' - q_2'$, and $p_2 =
2q_2' - q_1'$, while $p_i' = -\del H/\del q_i$ gives precisely the equations
of motion (\ref{toda}).   Thus the original equations of motion for the dyonic
black hole can be cast into the $SU(3)$ Toda equations (\ref{toda}).  The 
vanishing of the Hamiltonian is a consequence of requiring that the solution
be extremal.  In \cite{lpxtoda}, the general non-extremal case with 
non-vanishing
Hamiltonian was discussed.  We can obtain the extremal solution either by
taking the limit of the non-extremal solution in \cite{lpxtoda}, or by 
directly
solving the equations (\ref{toda},\ref{hamilton}).   
The required extremal solution
can be obtained by making the ansatz $e^{-q_2} = e^{-q_1} + {\rm const}$.
With this ansatz, it is easy to verify that $(e^{-q_1})'' = 1 = (e^{-q_2})''$.
Thus the solution is given by these two simple equations, subject to the first
order constraint (\ref{hamilton}).  Requiring that the function $A$ and
the dilaton $\phi$ be zero in the asymptotic limit $\rho = 1/r= 0$, 
the solution is
\bea
e^{\phi/\sqrt3 - 2A} &=& 1 + 4 Q_m^{2/3} (Q_e^{2/3} + Q_m^{2/3})^{1/2} 
\fft{1}{r} + 8 Q_e^{2/3} Q^{4/3}_m \fft{1}{r^2}\ ,\nonumber\\
e^{-\phi/\sqrt3 - 2A} &=& 1 + 4 Q_e^{2/3} (Q_e^{2/3} + Q_m^{2/3})^{1/2}
\fft{1}{r} + 8 Q_m^{2/3} Q_e^{4/3} \fft{1}{r^2}\ ,
\eea
together with $B=-A$.  There is an horizon at $r=0$.  All physical
quantities, including the dilaton, field strength and curvatures, are finite
on the horizon.  The entropy is 
\be
S = 8\pi Q_e Q_m\ ,
\ee
and the temperature $T$ is zero.  The non-extremal generalisation was given in
\cite{lpxtoda}.  It is easy to verify that in the near-extremal regime, the
entropy and temperature satisfy the relation
\be
S = 8 \pi Q_e Q_m + 128 (Q_e Q_m)^{4/3} \sqrt{Q_e^{2/3} + Q_m^{2/3}}\,
T\ .
\ee

      The mass of the dyonic black hole is 
\be
m= (Q_e^{2/3} + Q_m^{2/3})^{3/2} \ .\label{dyonmass2}
\ee
On the other hand, the total mass of the purely electric and magnetic
constituents, at large separation, is given by $m_\infty = Q_e + Q_m$.  Thus
the binding energy is negative whenever both charges are non-vanishing.

        It is interesting to note that if the two charges $Q_e$ and $Q_m$
are equal, $Q_e=Q_m = Q/\sqrt2$, the dilaton $\phi$ decouples, and the solution
becomes precisely the extremal $a=0$ Reissner-Nordstr\o{m} black hole, whose
metric is given by
\be
ds^2 = -\Big(1 + \fft{2Q}{r}\Big)^{-2} dt^2 + 
\Big(1 + \fft{2Q}r\Big)^2 dy^m dy^m\ ,
\ee
and its mass is $m=2Q$.  Thus we see that the $a=0$ Reissner-Nordstr\o{m}
black hole can be embedded in $D=4$ maximal supergravity in four
inequivalent ways.  One is the usual supersymmetric embedding with four
non-zero charges for appropriate field strengths, {\it e.g.}\ $F^{(12)}$,
$F^{(34)}$, $F^{(56)}$ and ${\cal F}^{(7)}$.\footnote{In fact this solution
itself has a non-supersymmetric variant \cite{lpmulti,ko}, achieved by making 
an alternative
sign choice for any one of the charges, which appear quadratically in the 
bosonic equations of motion. The fermionic equations of motion and
the supersymmetry transformations involve the field strength linearly, and thus
are sensitive to this sign choice.} In accordance with that fact that the 
solution is supersymmetric, this embedding has zero binding energy.  The other
three inequivalent embeddings are all non-supersymmetric, and involve
two charges, four charges or eight charges.  They have negative, zero and
positive binding energies respectively.  The first case is the one that
we have just discussed above.  The third case is the $a=0$ eight-charge black 
hole that we discussed in the previous section.  The second case, the 
$a=0$ non-supersymmetric
four-charge black hole with zero binding energy, can be obtained from the
dyonic black hole with two 2-forms that we discussed in the first paragraph 
of this section, by setting the charges $Q_1$ and $Q_2$ equal.

\section{Multi-scalar non-supersymmetric $p$-branes}

       In section 2 we discussed single-scalar $p$-brane solutions, in which
the non-vanishing charges occur in fixed ratios that are determined by the
dot products of the associated dilaton vectors $\vec a_\a$.  If the solution
is supersymmetric, it can easily be generalised to a multi-scalar solution
where all the charges become independent.  On the other hand, if the solution
is non-supersymmetric, such a generalisation is not known, except for the
new two-charge dyonic black hole which we discussed in the previous section. 
In this section, we shall present first-order perturbative solutions for 
charges
which are close to the fixed ratios of the exact single-scalar solutions. 
The Lagrangian is given by (\ref{dlag}). The equations of motion are
\cite{lpmulti}
\bea
&& \varphi_\a'' = \fft8{\td d^2} \epsilon \sum_\beta M_{\a\beta}\,Q_\b^2
\, e^{\epsilon \varphi_\beta + 2d A}\ ,\label{neqs}\\
&& d(D-2) A'^2 + \ft12 \td d \sum_{\a,\beta} (M^{-1})_{\a\beta} \varphi_\a'
\varphi_\beta' = \fft{8}{\td d} \sum_\a Q_\a^2 e^{\epsilon \varphi_\a
+ 2d A}\ ,\label{1storder}
\eea
where a prime again denotes a derivative with respect to $\rho = 1/r^{\td d}$,
$\varphi_\a = \vec a_\a \cdot \vec \phi$, and the function $A$ is given by
\be
A = \fft{\epsilon \td d}{D-2} \sum_{\a,\beta} (M^{-1})_{\a\beta} 
\, \varphi_\a\ .
\label{afunction}
\ee
Defining $\Phi_\a = \epsilon \sum_{\beta} (M^{-1})_{\a\beta}\, \varphi_\beta$,
it follows from (\ref{afunction}) that (\ref{neqs}) becomes
\be
\Phi_\a'' = \fft{8Q_\a^2}{\td d^2} \exp\Big({\sum_\beta (M_{\a\beta} + 
\fft{2d\td d}{ (D-2)}) \Phi_\beta}\Big)\ .\label{gentoda}
\ee
In the supersymmetric case, the dot products of dilaton vectors satisfy 
(\ref{susydot}), implying that these equations are diagonal Liouville
equations in $\Phi_\a$ and can be easily solved. The mass of the 
supersymmetric $p$-brane is given by
\be
m=\sum_\a Q_\a\ ,\label{susymass}
\ee
which is exactly the same as the total mass of the constituents when they 
are widely 
separated.  In general, the equations (\ref{gentoda}) have the general 
form of Toda equations, but with precise
coefficients that seem to render them non-integrable.  However, the exact 
solution can be obtained if the charges occur in the fixed ratios given by
(\ref{sscharge}) since then the equations reduce to those for the 
single-scalar solutions.  In this case, $\Phi_\a =\Phi_\a^0 \equiv
\epsilon c_\a^2
\phi/a$, where $\phi$ is the remaining dilaton of the single-scalar
solution, given by (\ref{ssol}).

	Let us consider a perturbation in which the charges $Q_\a$ are
displaced slightly from their single-scalar values:
\be
Q_\a = c_\a Q(1 + \ve_\a)\ ,\qquad
\Phi_\a = \Phi_\a^0 + \sum_\beta \ve_\b f_{\a\beta}\ .\label{perturb}
\ee
Substituting these equations into (\ref{gentoda}), we obtain the equations
of motion for the first-order functions $f_{\a\beta}$:
\be
f_{\a\beta}'' =\fft{8Q^2}{\td d^2} e^{\epsilon \Delta \phi/a} 
\Big(2 c_\a^2 \delta_{\a\beta} + a^2 \sum_{\gamma,\delta} 
(M^{-1})_{\a\delta} (M_{\a\gamma} + \fft{2d\td d}{D-2}) f_{\gamma\beta}\Big)
\ ,
\ee
where the only summations are those indicated explicitly.  Defining 
$f_\a = \sum_\beta f_{\beta\a}$, we have
\be
f_\a'' = \fft{8Q^2}{\td d^2} e^{\epsilon \Delta\phi/a} \Big(
2 c_\a^2 + \Delta f_\a\Big)\ .
\ee
It follows from (\ref{ssol}) that we can solve for the functions $f_\a$, 
obtaining\footnote{We have chosen the constants of integration so that the
perturbations $f_{\a}$, like the original functions $\Phi^0_\a$,
vanish at infinity ({\it i.e.}\ at $\rho=0$), and also so that the 
perturbations remain finite on the horizon $\rho=\infty$.}
\be
f_\a = \fft{2 c_\a^2}{\Delta } \Big((1+k\rho)^{-1} -1\Big)\ .
\ee
Thus we obtain the function $A$ in the metric, given by
\be
A = \fft{\td d}{D -2} \sum_\a \Phi_\a = A^0 + 
\fft{2\td d}{D-2} \sum_\a f_\a \varepsilon_\a\ ,
\ee
where $A^0$ is the unperturbed metric function.  Thus we have
\be
e^{2A}=\Big ( 1 + k\rho\Big)^{-\ft{4\td d}{\Delta (D-2)}}\, \Big( 1 +
\fft{4\td d}{(D-2)\Delta} ((1+ k\rho)^{-1} -1)\sum_\a c_\a^2 \ve_\a 
\Big)\ ,
\ee
to first order in $\ve_\a$,
where $\rho = r^{-\td d}$ and $k=2Q \sqrt\Delta/\td d$.  The mass per 
unit $p$-volume of the perturbed $p$-brane is 
\be
m=\fft{2Q}{\sqrt{\Delta}} \Big( 1 + \sum_\a c_\a^2 \ve_\a\Big)
\ .\label{genmass}
\ee

     It is easy to verify that the small perturbation mass formula 
(\ref{genmass}) is consistent with the general mass formula (\ref{susymass})
for supersymmetric solutions, since in this case, $c_\a^2 = 1/N$ and $\Delta
=4/N$.  It is also consistent with the mass formula (\ref{dyonmass2}) for
the new dyonic black hole in the previous section.  It would be of interest
to know the exact mass formulae for the cases that we have been analysing
perturbatively in this section.  At least for the cases where the $c_\a$ are 
all equal, a natural conjecture for the exact mass formula would be
\be
m = \Big(\sum_\a Q_\a^x\Big)^{1/x}\ ,\label{ecmass}
\ee
where $x$ is a constant that can be determined by requiring consistency 
with the mass for the single-scalar
solution where all the charges are equal.  Thus we have $x= \log N/\log
(2\sqrt{N/\Delta})$.  This relation correctly produces $x=1$ if we apply
it to the supersymmetric cases, and $x=2/3$ for the new black hole dyon
of section 3.   It is straightforward to show that (\ref{ecmass}) is also
consistent with the perturbative result (\ref{genmass}).  If one would be
able to generalise the non-supersymmetric equal-charge single-scalar 
solutions discussed in section 2 to multi-scalar solutions, it would 
follow from 
(\ref{ecmass}) that $x > 1$, and hence all these solutions would have
positive binding energy for all values of the charges.

\section{Conclusions}

      In this paper, we have argued that non-supersymmetric $p$-branes
can be viewed as bound states of $\Delta=4$ supersymmetric $p$-branes.  
In general the binding energy is non-zero, with a sign that depends on the
specific choice of the constituents, and in particular on the dot products 
$M_{\a\beta}$ of their dilaton vectors.  By contrast, the single-scalar 
supersymmetric 
$p$-branes with $\Delta=4/N = 2, 4/3, \ldots$ are also bound states of
appropriate $\Delta=4$ 
constituents, but with zero binding energy.  In fact it is not clear that
the single-scalar supersymmetric solutions can be sensibly interpreted
as distinct entities in their own right, since there is nothing (other
than initial conditions imposed to infinite precision) to prevent
the different charge species from drifting apart to give a multi-scalar
multi-center solution.  In other words, these supersymmetric single-scalar
$p$-branes are nothing but multi-scalar multi-center solutions where,
improbably, the centers happen to have become coincident.\footnote{
It is curious that the three-charge black hole in $D=5$ and four-charge
black hole in $D=4$ have non-vanishing entropy, which can be understood
microscopically from string theory \cite{sv,jkm,sm}, and yet an infinitesmial 
displacement of their constituents, at no cost in energy, would cause the 
classical entropy to vanish.} 

        The existence of non-supersymmetric solutions with positive binding
energy indicates that these configurations are energetically more favourable
than 
those where the associated constituents remain separated.  Thus although the
widely-separated constituents with the charge quantum numbers of a 
supersymmetric $p$-brane will be stable, if the charges are instead those
of a non-supersymmetric $p$-brane with positive binding energy, the 
constituents will be unstable to collapse.  Of course, the non-supersymmetric
$p$-brane may itself be unstable to quantum corrections.  Nontheless, the
existence of the lower-energy classical configuration is an indication of
the instability of the widely-separated constituents, each of which is 
asymptotically supersymmetric locally in its neighbourhood, even at
the quantum level.   On the other hand, non-supersymmetric $p$-branes with
negative binding energy indicate that the associated constituents in such
cases will tend to repel one another.

\section*{Note Added}

       After submitting this paper, we learned that the dyonic black hole
in section 3 was also discussed in \cite{gw,gk}, and its fission into
electric and magnetic black holes was discussed in \cite{gk}.

\end{document}